\documentstyle[12pt,amstex,amssymb,righttag]{article}
\begin{document}
\renewcommand{\thefootnote}{\fnsymbol{footnote}}

\thispagestyle{empty}

\vspace*{-1cm}
\begin{center}
{\Large \bf Non-Abelian Magnetic Monopoles in a Background of
Gravitation with Fermions}

\vspace{3mm}
by\\
\vspace{3mm}
{\sl Carlos
Pinheiro$^{\ddag}$\footnote{fcpnunes@@cce.ufes.br/maria@@gbl.com.br},
{\sl Cintia G. Pinheiro}$^{\ddag}$
} 
and 
{\sl F.C. Khanna$^{+}$\footnote{khanna@@phys.ualberta.ca}}

\vspace{3mm}
$^{\ddag}$Universidade Federal do Esp\'{\i}rito Santo, UFES.\\
Centro de Ci\^encias Exatas\\
Av. Fernando Ferrari s/n$^{\underline{0}}$\\
Campus da Goiabeiras 29060-900 Vit\'oria ES -- Brazil.\\

$^{+}$Theoretical Physics Institute, Dept. of Physics\\
University of Alberta,\\
Edmonton, AB T6G2J1, Canada\\
and\\
TRIUMF, 4004, Wesbrook Mall,\\
V6T2A3, Vancouver, BC, Canada.
\end{center}

\vspace{3mm}
\begin{center}
Abstract
\end{center}

The purpose of this paper is to study static solution configuration
which describes the magnetic monopoles in a scenary where the
gravitation is coupled with Higgs, Yang-Mills and fermions. We are
looking for analysis of the energy functional and Bogomol'nyi equations.
The Einstein equations now take into consideration the fermions'
contribution for energy-momentum tensor. The 
interesting aspect here is to verify that the fermion field gives a
contribution for non abelian magnetic field and for potential
which minimise the energy functional.

\newpage
\section{Introduction}
\setcounter{footnote}{0}

The study of magnetic monopoles started with the Dirac's 
paper in 1948 \cite{tres} where for the first time the possibility of
existence of this kind of particles in nature was analysed.

Some years later t'Hooft and Polyakov again found the magnetic
monopoles but in a different context, the so called non-abelian
monopoles had a scalar field as a source.

More recently Wali \cite{um,dois} has studied the Bogomol'nyi
equations and he again obtained the magnetic monopole type of
solution. The minimization of the functional of energy gives us the
conditions for analysis of Bogomol'nyi equations.

Finally Christiansen at all \cite{cinco} have studied the Bogomol'nyi
equations in Gauge theory and they verify conditions on functional of
energy which is suitable for obtaining finite energy for a static
configuration of field (gauge and scalar field) in $D=2+1$
dimensions.

In our case the objective is to study the
Einstein-gauge-field-Higgs-fermions system in $D=3+1$ dimensions and,
following  of Christiansen and Wali, we wish to
know the contribution for magnetic field and the potential that
minimizes 
the energy functional when we include fermions in the model. Second
we verify the possibility of the existence of non abelian magnetic
monopoles in $D=3+1$ dimension with fermion contributions.

The paper is written with the following outline:
\begin{enumerate}
\item
First we consider the complete system Higgs-gauge-fermions and
coupling between gravitation and scalar field. We have established 
equations of motion for systems with and without fermions.
\item
We review some results of Wali at al \cite{um,dois} for Bogomol'nyi
equations and for energy functional when fermions and gravitation are
not present.
\item
Next we consider the contribution of fermions alone but without
gravitation. There is no coupling between the scalar field and gravitation.
We obtain here the magnetic field with fermionic contributions and the
potential function that minizes the energy functional.
\item
We verify if there is a structure for the magentic monopole
when we consider the covariant divergence associated with the gauge field.
\item
Finally we return to the original problem and consider the
lagrangean with all fields and coupling between gravitation
and scalar field. All the equations are solved for this case, but
there remain two problems that have no solutions.

The first problem is about the new magnetic field and the potential
function when we account for the fermion contributions and the
background of gravitation.  

The questions is: what is the new magnetic field and potential
function that minimizes the functional of energy if we put fermions
and gravitation?

Second, what is the meaning of the presence of spinors in the initial
lagrangean? What is the physical interpretation for spinors 
here? 

In reality we have been speaking about ``fermions''. However in  a classical 
problem, the correct 
description is ``spinors''. The question is what is the meaning of the
spinors in our problem?

Anyway we have a partial solution for this problem and it will appear
elesewhere.
\end{enumerate}

Let us start with a non Abelian Higgs-gauge-``fermions'' and coupling
between scalar field and gravitation in $3+1$ dimensions. We would
like to verify the behaviour of magnetic monopole of 
t'Hooft-Polyakov type in a background of gravitation with presence of fermions.

We take the action given by Wali \cite{um} \cite{dois} for a
system of Einstein-Yang-Mills-Higgs in the form:
\begin{equation}
S=\int d^4x\sqrt{-g}\ 
\left[\frac{1}{16\pi Gv^2}\ R\phi^2-\frac{1}{4}
\left(F^a_{\mu \nu}\right)^2+\frac{1}{2}\left(D_{\mu}\phi^a\right)^2
-\frac{\lambda}{4}\left(\phi^2-v^2\right)^2\right]
\end{equation}
and now we will consider the Dirac Lagrangean
\begin{equation}
{\cal L}=\overline{\psi}\mbox{}^i_{\alpha}\left(i\gamma^{\mu}_{\alpha \beta}
{\cal D}_{\mu}-m\delta_{\alpha \beta}\right)\psi^i_{\beta}+\chi \overline{\psi}\mbox{}^i_{\alpha}
\phi^a\left(T^a\right)_{ij}\psi^j_{\alpha}
\end{equation}
for fermion contribution.

The matrix $T^a$ are hermitian with null trace. It describes the three
generators of $SU(2)$ group, here $a=1,2,3$ in the $N$-dimensional
representation. The indices $i,\ j=1,\cdots N$, where $N$ represent
the dimension of given irreducible representation of $SU(2)$. The
$\chi$ is only a constant. 

The signature of metric is given by $(+ - - -)$. The indices $\mu
,\nu$ take values from $\underline{0}$ to $\underline{3}$ and the 
indices $i,\ j$  assume values between $1$ and $3$. The indices $a,b$
vary with the representation of a given gauge group.

We define $R$ as the scalar curvature, $F_{\mu \nu}\mbox{}^a$
represents the stress field associated with the gauge field
$A^a_{\mu}$ given as
\begin{equation}
F^a_{\mu \nu}=\partial_{\mu}A^a_{\nu}-\partial_{\nu}A^a_{\mu}+gf^{abc}A^b_{\mu}
A^c_{\nu}
\end{equation}
where $f^{abc}$ represent the structure constants of the group. The
gauge covariant derivative associated with Higgs field $\phi^a$ is
given by
\begin{equation}
D_{\mu}\phi^a=\partial_{\mu}\phi^a+gf^{abc}A^b_{\mu}\phi^c\ .
\end{equation}
A new covariant derivative associated with the spinor field is necessary
here. The fermion field derivative $\psi^i_{\beta}$ is given by 
\begin{equation}
{\cal D}_{\mu}\psi^i_{\beta}=\partial_{\mu}\psi^i_{\beta}-igA^a_{\mu}
\left(T^a\right)_{ij}\psi^j_{\beta}
\end{equation}
where $g$ is the coupling constant.

The term $
\displaystyle{\frac{\lambda}{4}}\left(\phi^2-v^2\right)^2$ represents
the possibility of gauge symmetry breaking. We choose the unit system
such that $4\pi Gv^2=1$. Thus the action is written as
\begin{eqnarray}
S &=& \int d^4x\sqrt{-g}\Big[-\ \frac{~1~}{4}\ R\phi^2-\ \frac{~1~}{4}\ 
\left(F_{\mu \nu}\mbox{}^{a}\right)+\frac{1}{2}
\left(D_{\mu}\phi^a\right)^2+U\left(\phi_i\right)+\\
&{}&\overline{\psi}\mbox{}^i_{\alpha}\left(i\gamma^{\mu}_{\alpha \beta}{\cal D}_{\mu}-
m\delta_{\alpha \beta}\right)\psi^i_{\beta}+\chi \overline{\psi}\mbox{}^i_{\alpha}\phi^a
\left(T^a\right)_{ij}\psi^j_{\alpha}\Big]\ . \nonumber 
\end{eqnarray}
Here we use $U\left(\phi_i\right)$; $i=1,2,3$ in the form
given by (1.1).

The equations of motions for Yang-Mills and Higgs fields when fermion
are not considered are given respectively by
\begin{eqnarray}
&{}&\frac{1}{\sqrt{-g}}\ D_{\mu}\left(\sqrt{-g}\ F^{\mu \nu a}\right)=-
gf^{abc}\phi^bD^{\nu}\phi^c\ ,\\
&{}&\frac{1}{\sqrt{-g}}\ D_{\mu}\left(\sqrt{-g}\ D^{\mu}\phi^a\right)=-
\left[\frac{~R~}{2}+ \lambda \left(\phi^2-v^2\right)\right]\phi^a
\end{eqnarray}
and the Einstein equations are given by
\begin{equation}
G_{\mu \nu}=R_{\mu \nu}-\frac{1}{2}\ g_{\mu \nu}R=
\frac{2}{\phi^2}\ T_{\mu \nu}
\end{equation}
where the stress energy-momentum tensor for Einstein Yang-Mills-Higgs
is written as
\begin{equation}
T_{\mu \nu}=g_{\mu \nu}T-F_{\mu \rho}\mbox{}^{a}F_{\nu}\mbox{}^{\rho a}+
D_{\mu}\phi^aD_{\nu}\phi^a-\frac{1}{2}\ \phi^2
\left(R_{\mu \nu}-\frac{1}{2}\ g_{\mu \nu}R\right)
\end{equation}
and
\begin{equation}
T=\frac{1}{4}\ F_{\mu \nu}\mbox{}^{a}F^{\mu \nu a}=\frac{1}{2}\ 
D_{\mu}\phi^aD_{\nu}\phi^a+\frac{\lambda}{4}
\left(\phi^2-v^2\right)^2
\end{equation}
The contribution to the energy-momentum due the fermions is given by 
\begin{eqnarray}
T_{\mu \nu} &=& \frac{1}{2}\ i\left(\overline{\psi}\mbox{}^i_{\alpha}\gamma_{\nu
\alpha \beta}{\cal D}_{\mu}\psi^i_{\beta}+\overline{\psi}\mbox{}^i_{\alpha}
\gamma_{\mu \alpha \beta}{\cal D}_{\nu}\psi^i_{\beta}\right)+
g_{\mu \nu}\left(m\overline{\psi}\mbox{}^i_{\alpha}\psi^i_{\beta}\delta^{\alpha \beta}+
\right.\nonumber \\
&-& \left.ie^{\chi}_{a}\psi^i_{\alpha}\gamma^a_{\alpha \beta}{\cal D}_{\chi}\psi^i_{\beta}-
\chi\overline{\psi}\mbox{}^i_{\alpha}\phi^a\left(T^a\right)_{ij}\psi^i_{\alpha}\right)
\end{eqnarray}
Now with fermions, the equations of motion for fields 
$A^a_{\mu}$  $\phi^a$ and $\psi^i_{\beta}$ are given respectively by 
\begin{eqnarray}
&{}& \partial_{\beta}F_{\alpha \beta}\mbox{}^{d}=gf^{acd}
\Big[F_{\alpha \nu}\mbox{}^{a}A^c_{\nu}-\left(D_{\alpha}\phi^a\right)
\phi^c\Big]+g\psi^i_{\varepsilon}\gamma^{\mu}_{\varepsilon \gamma}
\delta_{\mu \alpha}\left(T^d\right)_{ij}\psi^j_{\gamma}\
,\nonumber \\
&{}&\left(-\frac{1}{\sqrt{-g}}\ \partial_{\mu}\sqrt{-g}\right)D_{\mu}\phi^a=
D_{\mu}\left(D_{\mu} \phi^a\right)+\left[\frac{R}{2}+\lambda 
\left(\phi^2-v^2\right)\right]\phi^a-\chi \overline{\psi}\mbox{}^i_{\alpha}
\left(T^a\right)_{ij}\psi^j_{\alpha}\ , \nonumber\\
&{}&\\
&{}&-\ \frac{1}{\sqrt{-g}}\left(\partial_{\mu}\sqrt{-g}\right)\bar{\psi}^k_{\alpha}
i\gamma^{\mu}_{\alpha \gamma}=\gamma^{\mu}_{\alpha \gamma}{\cal D}_{\mu}
\bar{\psi}^k_{\alpha}+m\bar{\psi}^i_{\alpha}\delta^{\gamma}_{\alpha}
\delta^i_k-\chi \bar{\psi}^i_{\gamma}\phi^a\left(T^a\right)_{ik}\ ,
\end{eqnarray}

\section{Static Equations and Bogomol'nyi Conditions}\setcounter{equation}{0}

\paragraph*{}
We wish to get only static solutions (time independent) for the system
described by (1.6). Thus,  using the technique of Bogomol'nyi, with
appropriate boundery conditions \cite{um,dois,cinco},
\begin{equation}
D\phi =0\quad , \quad \phi^2=v^2\ .
\end{equation}
The gauge field without fermions is given by \cite{cinco}
\begin{equation}
A^a_i=\frac{1}{gv^2}\ \varepsilon^{abc}\phi^b\partial_i\phi^c+
\frac{1}{v}\ \phi^aA_i
\end{equation}
where $A_i$ is arbitrary and $F_{ij}\mbox{}^{a}$ satisfy 
\begin{equation}
\phi^aF_{ij}\mbox{}^{a}=\frac{\phi^2}{v}\ {\cal F}_{ij}
\end{equation}
and the field ${\cal F}_{ij}$ is given by
\begin{equation}
{\cal F}_{ij}=\frac{1}{gv\phi^2}\ \varepsilon^{abc}\phi^a\partial_i\phi^b\partial_j
\phi^c+\partial_iA_j-\partial_jA_i\ .
\end{equation}
Only the static abelian gauge field will survive for long distances.
Then, we define the ``magnetic field'' ${\cal B}_i$ associated with the
monopole for long distance as 
\begin{equation}
{\cal B}^i=\frac{1}{2}\ \varepsilon^{abc}{\cal F}_{jk}=\frac{1}{2gv^3}\ 
\varepsilon^{ijk}\varepsilon^{abc}\phi^a\partial_j\phi^b\partial_k\phi^c+
\varepsilon^{ijk}\partial_jA_k\ .
\end{equation}
The magnetic charge of the configuration is given by 
\begin{equation}
g=\frac{1}{4\pi}\int d^3x\partial_i{\cal B}^i=\frac{1}{8\pi gv^3}\int_{s^2_{\infty}}
d\sigma_i\varepsilon^{ijk}\varepsilon^{abc}\phi^a\partial_j\phi^b\partial_k
\phi^c=\frac{n}{g}
\end{equation}
where
\begin{equation}
n=\frac{1}{8\pi v^3}\int_{s^2_{\infty}}d\sigma_i\varepsilon^{ijk}
\varepsilon^{abc}\phi^a\partial_j\phi^b\partial_k\phi^c\ .
\end{equation}
which is a topological number.

Taking the limit
$\displaystyle{\frac{\lambda}{g}\rightarrow 0}$, the functional of
energy obtained from (1.6), without fermions for the flat spacetime is
given by
\begin{equation}
\varepsilon =\int d^3x\left(\frac{1}{4}\ F_{ij}\mbox{}^{a}F^{ija}-
\ \frac{1}{2}\ D_i\phi^aD^i\phi^a\right)
\end{equation}
that is an energy of Higgs-Yang-Mills field. 

For the case with the fermions, we have the energy functional
written as,
\begin{eqnarray}
\varepsilon &=&\int d^3x\Big[\frac{1}{4}\ F_{ij}\mbox{}^{a}F^{ija}-
\frac{1}{2}\ D_i\phi^aD^i\phi^a+\overline{\psi}\mbox{}^i_{\alpha}
\left(m\delta_{\alpha \beta}-i\gamma^k_{\alpha \beta}{\cal D}_k\right)
\psi^i_{\beta}+\nonumber \\
&-&\chi \overline{\psi}\mbox{}^i_{\alpha}\phi^a\left(T^a\right)_{ij}
\psi^j_{\alpha}+U\left(\phi_i\right)\Big]\ .
\end{eqnarray}

We are not considering the coupling between the scalar and gravitation here. 

>From eq. (2.8) we can define the electric and magnetic fields as:
\begin{eqnarray}
&{}&E_i\mbox{}^{a}= D_i\phi^a\ , \\
&{}& B_i^a = \frac{1}{2}\ \varepsilon_{ijk}F^{jka}
\end{eqnarray}
for the case when the fermions are not present and when the flat
spacetime is considered $(R=0)$; in other words, the eq.
(2.10) and (2.11) are good definitions for the case 
when we don't consider the gravitation background. If we use the radiation
gauge, $A^a_0=0$, and for  the static solution $D_0\phi^a=0$, from eq. (1.3) and (1.4) it follows
that 
\begin{eqnarray}
F_{oi}\mbox{}^{a} &=&
\partial_oA^a_i-\partial_iA^a_o+gf^{abc}A_o^bA^c_i\ ,
\nonumber \\
&{}&F_{oi}\mbox{}^{a}=0=D_i\phi^a\ .
\end{eqnarray}

Then we have obtained a solution (null electricaly), because the
electric field is zero in the whole space. 

We have still \cite{um,dois} the following inequality.
\begin{eqnarray}
\varepsilon &=& \frac{1}{2}\int d^3x\left[\left(E^a_i\right)^2+
\left(B_i^a\right)^2\right]^2\nonumber \\
&=& \frac{1}{2}\int d^3x\left(E^a_i\mp B^a_i\right)
\left(E^a_i\mp B^a_i\right)\pm \int d^3xE_i^aB^a_i\geq \pm \int
d^3xE^a_iB^a_i\ .
\end{eqnarray}
Using now the Bianchi indetity for $F_{ij}\mbox{}^{a}$, we can write 
\begin{equation}
\pm \int d^3xE_i^aB^a_i=\pm \int d^3x\partial_i\left(
\frac{1}{2}\ \varepsilon^{ijk}F_{jk}^a\phi^a\right)
\end{equation}
Comparing eq. (2.14) with eq. (2.3) the surface integral in eq. (2.14)
asymptotically takes the value
\begin{equation}
\pm v \int d^3x\partial_iB^i=\frac{4\pi nv}{g}\ .
\end{equation}

On the other hand, if the Bogomol'nyi equations
\begin{equation}
E_i^a=B_i^a
\end{equation}
are satisfied, then the functional of energy is definitely 
minimized. 
It was shown in \cite{um} \cite{dois}, that this follows naturely from the
relations (2.13) -- (2.16).

Now, consider the case described by (2.9) where the fermions are
present but without the background of gravitation.

We can still write suitably the energy functional (2.9) in terms of
fields $E_i^a$ and $B^a_i$ with the same arguments. However, here it
will not have the electric component  due to the
presence of the fermions. The magnetic field will be different since 
it shall have the contribution of fermions.

The scalars fields will be treated as a condensate of
fermions. We shall define the following quantities:
\begin{eqnarray}
\eta &=& m\overline{\psi}\mbox{}^i_{\alpha}\psi^i_{\beta}\delta_{\alpha
\beta}\ ,\nonumber \\
\xi &=& i\overline{\psi}\mbox{}^i_{\alpha}\gamma_{\alpha \beta}^k\
{\cal D}_k\psi^i_{\beta}\ , \\
\hspace*{-4cm}\!\!\!\!\!\!\!\!\!\!\!\!\!\!\!\!\!\!\!\! \mbox{and}\  \hspace*{5cm}\Delta &=& \chi
\overline{\psi}_{\alpha}\mbox{}^i\phi^a\left(T^a\right)_{ij}\psi^j_{\alpha}\
.\nonumber
\end{eqnarray}
We have created three new scalar fields $(\eta ,\xi ,\Delta )$. So,
our Lagrangean (1.6) with the scalar curvature $R=0$ or its form
eq. (2.8) with the fermions present, it is not possible anymore to
obtain an 
energy functional that is satured by $\lambda \phi^4$ potential as in
\cite{cinco}. 

Using now the same prescription as in \cite{cinco} it is
conjectured that it is possible to verify that eq. (2.9) in
$D=3+1$ dimensions may be reduced to the following form:
\begin{eqnarray}
&\varepsilon =\displaystyle{\frac{ev^2}{2}}\ \Phi_{{}_{B^a_k}}\pm 
\displaystyle{\frac{1}{2e}}\oint d\sigma_i
J^a_j\varepsilon^{ijk}\varepsilon^{abc}\phi^b\partial_k\phi^c\mp \int
d^3x\Big\{\displaystyle{\frac{1}{2}}\left(B^a_k\partial_k\phi^a\mp
\sqrt{2U}\right)^2 &\nonumber \\
&\!\!\!\mp \displaystyle{\int} \left[\mp \displaystyle{\frac{e}{2}}\left(v^2-|\phi |^2\right)
\left(\eta +\xi +\Delta \right)\left(\partial_k\phi^a\right)\mp 
\sqrt{2U}\right]B^a_k\partial_k\phi^a+\displaystyle{\frac{1}{2}}\left|\left(
D_1\pm iD_2\right)\phi^a\right|^2\Big\}{~}\!\! .&
\end{eqnarray}
Here $\Phi_{{}_{B^a_k}}$ means the non-abelian magnetic flux. The second
term is a surface term and goes to zero at infinity since all
fields go to zero at infinity.

We need to discover what magnetic field $B^a_k$ and potential
$U(\phi_i)$  will minimize the energy functional or in other
words, what are $B^a_k$ and $U(\phi_i)$ which will saturate the
functional, $\varepsilon$?

If we use the duality condition \cite{cinco}  as in the form 
\begin{equation}
D_1\phi^a=\mp iD_2\phi^a
\end{equation}
and noting that we are considering only configurations with finite
energy the surface integral of current vector is null.

In the Bogolomol'nyi limit given by eq. (2.18) the minimum of
energy is obtained exactly if
\begin{equation}
\varepsilon =\frac{ev^2}{2}\ \Phi_{{}_{B^a_k}}\ .
\end{equation}

Since we wish to saturate the functional $\varepsilon$ the magnetic
field now carrying the fermion's contribution will be given by 
\begin{equation}
{\cal B}^a_k=\frac{e}{2}\left(v^2-|\phi |^2\right)\left(\eta +\xi +\delta \right)
\left(\partial_k\phi^a\right)\ .
\end{equation}
and the new potential will be written as
\begin{equation}
U\left(\phi^a,\eta ,\xi ,\Delta \right)=\frac{e}{8}
\left(|\phi |^2-v^2\right)^2\left(\eta +\xi +\Delta \right)^2
\left(\partial_k\phi^a\right)^4
\end{equation}
Clearly the potential is now of type $\lambda \phi^8$ because of
fermion contribution.

With the definition of the covariant derivative given by
\[
D_{\mu}=\partial_{\mu}+i\tilde{g}A_{\mu}
\]
we get 
\[
D_{\mu}F^{\mu \nu a}=\partial_{\mu}F^{\mu \nu a}+i\tilde{g}
\left[A_{\mu},F^{\mu \nu}\right]^a\ .
\]
Thus, the covariant divergence is given as
\begin{equation}
D_{\mu}F^{\mu \nu a}=\vec{\nabla}\cdot B^a+\tilde{g}f^{abc}\vec{A}_b
\cdot \vec{B}_c=0
\end{equation}
or still in a compact form,
\begin{equation}
D_{\mu}F^{\mu \nu a}=\vec{D}\cdot \vec{B}=0
\end{equation}
where
\begin{equation}
D=\vec{\nabla}+\tilde{g}f^{abc}\vec{A}\ .
\end{equation}

Since $B^a_k$ is given by eq. (2.21) and $|\phi |^2$ is written as 
\begin{equation}
|\phi |^2=\phi^a\phi_a 
\end{equation}
a suitably ansatz for scalar field may be chosen such as 
\begin{equation}
\phi^a=F\ \frac{r^a}{r}\quad (r\rightarrow \infty )\ .
\end{equation}
Using eq. (2.27) and eq. (2.26) in equation eq. (2.23) or eq. (2.24) above we get 
\begin{equation}
\vec{\nabla}\cdot \vec{B}^a_k\sim \frac{g}{r^2}\ .
\end{equation}
This  gives us a structure of the magnetic monopole
where $g$ represents the source of the magnetic field (non-Abelian
magnetic field) whose source has the origin in the scalar field, gauge
field and condensate of fermions 
$\left(\phi^a,A^a_{\mu},\eta ,\xi ,\Delta \right)$. 
It's sufficient to choose or to fix the ``$F$'' function such as that
form to
obtain the Gauss law from eq. (2.28).

\section*{The energy functional in a curved spacetime} 

\paragraph*{}
Now the same problem is proposed in a curved spacetime.

The energy functional (static functional) that is obtained from eq. (1.6)
in a curved spacetime is described by.
\begin{eqnarray}
\varepsilon &=& \int d^3x\sqrt{-g}\Big[\frac{1}{4}\ R\phi^2+\frac{1}{4}\ 
F_{ij}\mbox{}^{a}F^{ija}-\frac{1}{2}\ D_i\phi^aD^i\phi^a+
\frac{\lambda}{4}\left(\phi^2-v^2\right)^2 +\nonumber \\
&{}&\overline{\psi}\mbox{}^i_{\alpha}\left(m\delta_{\alpha \beta}-
i\gamma^k_{\alpha \beta}{\cal D}_k\right)\psi^i_{\beta}-
\chi
\overline{\psi}\mbox{}^i_{\alpha}\phi^a\left(T^a\right)_{ij}\psi^j_{\alpha}\Big]
\ .
\end{eqnarray}

The energy functional can, in principle, be  minimized with the gravitation as a
background field only if we introduce a third covariante derivative
associated with gravitation, but we need also to transfer the
dynamics 
from the metric to vierbein and to write an appropriate covariante
derivative carrying the spin connection such as:
\[
\tilde{D}_{\mu}=\partial_{\mu}+\frac{1}{8}\ i \left[\gamma_a\gamma_b\right]
B_{\mu}^{ab}
\]
where 
\[
\gamma^{\mu}=e^{\mu}_a(x)\gamma^a
\]
and 
\[
g_{\mu \nu}=e^a_{\mu}e^b_{\mu}\eta_{ab}
\]
are Dirac's matrices and metric respectively written in a local
Lorentz coordinate system, $e^a_{\mu}$ are the vierbeins and
$B^{ab}_{\mu}(x)$ are spin connections, completely determined by
vierbeins. Now the gravitation is considered under flat space-time
given by $\eta_{ab}$ in Minkowski space.

\section{The basic Equations}\setcounter{equation}{0}

\paragraph*{}
We choose the  spherically symmetric static metric as
\begin{equation}
ds^2=A^2(r)dt^2-B^2(r)dr^2-C^2(r)r^2d\Omega^2\ .
\end{equation}
The Einstein tensor and the scalar curvature are easily obtained,
\begin{eqnarray}
G_{00} &=& \frac{A^2}{B^2}\left[\frac{1}{r^2}\left(\frac{B^2}{C^2}-1\right)
+\frac{2}{r}\ \frac{B'}{B}+2\ \frac{B'}{B}\ \frac{C'}{C}-
\frac{2C''}{C}-\frac{6C'}{rC}-\left(\frac{C'}{C}\right)\right]\ , \nonumber
\\
G_{rr} &=& \left(\frac{C'}{C}\right)^2+\frac{2}{r}\ \frac{C'}{C}+\frac{1}{r^2}
\left(1-\frac{B^2}{C^2}\right)+\frac{2C'}{C}\ \frac{A'}{A}+\frac{2}{r}
\frac{A'}{A}\ , \\
G_{\theta \theta} &=& \frac{C^2r^2}{B^2}\left(\frac{A''}{A}+
\frac{1}{r}\ \frac{A'}{A}-\frac{A'}{A}\ \frac{B'}{B}+\frac{C'}{C}\ 
\frac{A'}{A}-\frac{1}{r}\ \frac{B'}{B}-\frac{B'}{B}\ \frac{C'}{C}+
\frac{C''}{C}+\frac{2}{r}\ \frac{C'}{C}\right)\ ,\nonumber \\
G_{\phi \phi} &=& sen^2\theta G_{\theta \theta}\ ,\nonumber \\
R &=& \frac{2}{r^2}\left(\frac{1}{B^2}-\frac{1}{C^2}\right)+
\frac{2}{B^2}\left[\frac{A''}{A}+\frac{2}{r}\ \frac{A'}{A}-
\frac{B'}{B}\ \frac{A'}{A}-\frac{2}{r}\ \frac{B'}{B}-
\frac{2B'}{B}\ \frac{C'}{C}+\right.\nonumber \\
&{}&\left.\frac{2C'}{C}\ \frac{A'}{A}+\frac{2C''}{C}+\frac{6C'}{rC}+
\frac{C'}{C}\right]
\end{eqnarray}
Then if we consider the eq. (1.10) and (1.11) together with the
eq. (3.2) we take components of the energy momentum tensor for 
the global system of Einstein-gauge-Higgs-fermions. 

The components are given by:
\begin{eqnarray}
\tilde{T}_{00} &=& \frac{\phi^2A^2}{B^2}\left[
\frac{2C'}{C}\ \frac{A'}{A}+\frac{2}{r}\ \frac{A'}{A}-\frac{A'}{A}\
\frac{B'}{B}+\frac{A''}{A}-\frac{1}{2}\left(\frac{A'}{A}\right)^2+
\frac{\lambda v^2}{4}\ \frac{\left(h^2-1\right)^2}{h^2}+\right.\nonumber \\
&{}& \left.\frac{2h''}{h}+\left(\frac{2h'}{h}\right)+\frac{2h'}{h}\left(
\frac{2C'}{C}+\frac{2}{r}-\frac{B'}{B}\right)+\frac{3A'}{A}\ 
\frac{h'}{h}\right]+\tau_{00}\ , \nonumber \\
\tilde{T}_{rr} &=& \phi^2\left[-\frac{1}{2}\left(\frac{A'}{A}\right)^2-
\frac{A'}{A}\ \frac{h'}{h}+\frac{2B'}{B}\ \frac{h'}{h}-
\frac{\lambda v^2}{4}\
B^2\frac{\left(h^2-1\right)^2}{h^2}\right]+\tau_{11}\ ,\nonumber \\
\tilde{T}_{\theta \theta} &=& \frac{C^2r^2}{B^2}\ \phi^2\left[\frac{1}{2}\left(
\frac{A'}{A}\right)^2-\frac{h''}{h}-\left(\frac{h'}{h}\right)^2-\frac{h'}{h}
\left(\frac{2C'}{C}+\frac{2}{r}-\frac{B'}{B}\right)\right.+\nonumber \\
&{}&\left.-\frac{\lambda v^2}{4}\ \frac{B^2\left(h^2-1\right)^2}{h^2}\right]+
\tau_{22}\ , \\
\tilde{T}_{\phi \phi} &=& sen^2\theta \tilde{T}_{\theta
\theta}+\tau_{33}\ \nonumber 
\end{eqnarray}
where $h=h(r)=\displaystyle{\frac{\phi^a}{vr^a}}$ and $\tau_{00},\
\tau_{11},\ \tau_{22}, \ \tau_{33}$ are the components of the  
energy-momentum tensor for fermions given by
\begin{eqnarray}
\tau_{00} &=& i\overline{\psi}\mbox{}^i_{\alpha}
\gamma_{0\alpha \beta}{\cal D}_0\psi^i_{\beta}+g_{00}
\left(m\overline{\psi}\mbox{}^i_{\alpha}\psi^i_{\beta}\delta^{\alpha}_{\beta}
-ie^{\chi}_a\psi^i_{\alpha}
\gamma^a_{\alpha \beta}{\cal D}_{\chi }\psi^i_{\beta}
-\chi \psi^i_{\alpha}\phi^a\left(T^a\right)_{ij}
\psi^j_{\alpha}\right)\ , \nonumber \\
\tau_{11} &=& i\overline{\psi}\mbox{}^i_{\alpha}
\gamma_{1\alpha \beta}{\cal D}_1\psi^i_{\beta}+g_{11}
\left(m\overline{\psi}\mbox{}^i_{\alpha}\psi^i_{\beta}
\delta^{\alpha}_{\beta}-ie^{\chi}_a\psi^i_{\alpha}\gamma^a_{\alpha \beta}
{\cal D}_{\chi}\psi^i_{\beta}-
\chi \overline{\psi}^i_{\alpha}\phi^a\left(T^a\right)_{ij}
\psi^j_{\alpha}\right)\ , \nonumber \\
\tau_{22} &=& i\overline{\psi}\mbox{}^i_{\alpha}
\gamma_{2\alpha \beta}{\cal D}_2\psi^i_{\beta}+g_{22}
\left(m\overline{\psi}\mbox{}^i_{\alpha}\psi^i_{\beta}
\delta^{\alpha}_{\beta}-ie^{\chi}_a\psi^i_{\alpha}\gamma^a_{\alpha \beta}{\cal D}_{\chi}\psi^i_{\beta}-
\chi \overline{\psi}\mbox{}^i_{\alpha}\phi^a\left(T^a\right)_{ij}
\psi^j_{\alpha}\right)\ , \nonumber \\
\tau_{33} &=& i\overline{\psi}\mbox{}^i_{\alpha}
\gamma_{3\alpha \beta}{\cal D}_3\psi^i_{\beta}+g_{33}
\left(m\overline{\psi}^i_{\alpha}\psi^i_{\beta}
\delta^{\alpha}_{\beta}-ie^{\chi}_a\psi^i_{\alpha}\gamma^a_{\alpha \beta}
{\cal D}_{\chi}\psi^i_{\beta}-
\chi \overline{\psi}\mbox{}^i_{\alpha}\phi^a\left(T^a\right)_{ij}
\psi^j_{\alpha}\right)
\end{eqnarray}
where $e^{\chi}_a$ imply the tetrads or vierbeins.

This set of equations together with the Bogomol'nyi equations are the
basic equations for our system, Einstein-gauge-Higgs-fermions. 

Now the next step is to find solutions for the Einstein equations
with the objective to find the monopoles appearing in this
gravitational background when the fermions are present. But on the
other hand when we have interaction between spinors and gravitation
field 
the unique way to consider that coupling in our case is only if we
introduce a local Lorentz coordinate system. Thus, we do not have curved
space-time anymore.

\section*{Conclusions}

\paragraph*{}
We have analysed two types of systems. One of them with Yang-Mills
and scalar fields in flat spacetime and other consisting of the
Yang-Mills-scalar field and ``fermions''. 

We treated ``the fermions'' as a condensate of scalars fields. In this
case the magnetic field that saturated the energy functional has contributions
from fermions. The potential which minimized the same functional is
of the kind $\lambda \phi^8$ and thus different from \cite{cinco}
whose potential is of type $\lambda \phi^4$ for $2+1$ dimension case.

For the case of Yang-Mills-scalar field \cite{um,dois} the
Bogomol'nyi equations have a simple solution in flat space time. In
the present 
case our conjecture in eq. (2.18) gives us the non abelian magnetic field
much more complicated and potential with fermion's contribution. A
structure of magnetic monopoles can be found with a field given by
eq. (2.28).

Finally we have assumed the system Einstein-Higgs-scalar field-gauge
and we have obtained all the equations for this system but 
solutions to the Einstein equations are lacking. Magnetic monopoles
appear in a new context and it's still necessary to find in a new
magnetic field $B^a_k$ and potential $U(\phi_i)$ that will be
suitable for minimising the functional of energy when the
background gravitational  field is considered. At this point we have
a conflict between a curve space-time and a local lorentz manifold. On
the first way we have the the complete set of equations plus
Bogolmo'ni equations being the basic equations for
Einstein-gauge-Higgs-fermions system- and, in principle, the energy
functional could be minimized with the graviation field as a
background, but the other hand because the spinors we need to
introduce a local coordinates system and to proceed the minimization
of the functional of energy via vierbeins and spin connection.
 This will be a part of
a separate analysis.

\subsection*{Acknowledgements:}

\paragraph*{}
I would like to thank the Department of Physics, University of
Alberta for their hospitality. This work was supported by CNPq
(Governamental Brazilian Agencie for Research.

I would like to thank also Dr. Don N. Page for his kindness and attention
with  me at Univertsity of Alberta.

\end{document}